\documentclass[fleqn,10pt]{wlscirep}
\usepackage[utf8]{inputenc}
\usepackage[T1]{fontenc}
\usepackage{lineno}
\usepackage{gensymb}
\usepackage{textcomp}
  \usepackage[flushleft]{threeparttable}
\usepackage{hyperref}
\usepackage{amsmath}

\title{Emission inventory for maritime shipping emissions in the North and Baltic Sea (2015)} 

\author[1, *]{Franziska Dettner}
\author[1]{Simon Hilpert}
\author[2]{Dr. Ronny Petrik}
\author[3]{Rolf Nagel}
\affil[1]{Europa-Universität Flensburg, Centre for Sustainable Energy Systems, Flensburg, 24943, Germany}
\affil[2]{Helmholtz-Zentrum Hereon, Max-Planck-Straße 1, 21502 Geesthacht}
\affil[3]{FSG-Nobiskrug Design GmbH, Batteriestraße 52, 24939 Flensburg}

\affil[*]{corresponding author: Franziska Dettner (franziska.dettner@uni-flensburg.de)}

\begin{abstract} 
A high temporal and spatial resolution emission inventory for the North Sea and Baltic Sea for the year 2015 was compiled using current emission factors and ship activity data. The inventory includes seagoing vessels over 100 GT registered with the International Maritime Organisation traversing in the North and Baltic Seas. A bottom-up approach was chosen for the compilation of the inventory, which provides emission levels of the air pollutants CO\textsubscript{2}, NO\textsubscript{x}, SO\textsubscript{2}, PM\textsubscript{2.5}, CO, BC, Ash, NMVOC and POA, as well as the speed-dependent fuel and energy consumption. Input data come from both main and auxiliary engines as well as \textit{well-to-tank} and \textit{tank-to-propeller} emission and energy and fuel consumption quantities. The geo-referenced data are provided in a temporal resolution of five minutes. The data can be used to assess, inter alia, the health effects of maritime emissions, external and social costs of maritime transport, emission mitigation effects of alternative fuel scenarios and shore-to-ship power supply. 
\end{abstract}
\begin{document}

\flushbottom
\maketitle

\thispagestyle{empty}


\section*{Background \& Summary} 
Approximately 3\% of global carbon dioxide (CO\textsubscript{2}) emissions and a substantial amount of other harmful air pollutant emissions originate from international maritime shipping \cite{imo_fourth_2020}. Maritime transport is expected to increase significantly in the coming years and according to the International Maritime Organisation (IMO), maritime CO\textsubscript{2} emissions could increase by 50-250\% by 2050 under a business-as-usual scenario \cite{imo_fourth_2020}. If mitigation measures are not imposed, this is likely to undermine the 1.5\textdegree target set by the Paris Agreement. Ship emissions of CO\textsubscript{2}, nitrogen oxides (NO\textsubscript{x}) and particulate matter (PM) must therefore be regulated by directives in order to reduce harm to the climate and health. For the analysis of guidelines and emission limits, as well as the climate and health effects of shipping emissions, a comprehensive and transparent maritime emissions inventory is essential.

Before 2004, an estimate of maritime emissions was only possible by estimating fuel consumption through the amount of bunkered fuel oil. Initial studies were carried out by Corbett \textit{et al.}\cite{corbett_emissions_1997}, \cite{corbett_global_1999}, who quantified nitrogen and sulphur emissions from global shipping for 1993, and Eyring \textit{et al.}\cite{eyring_emissions_2005-1}, who calculated CO\textsubscript{2} emissions from international shipping to be 187 Tg in 1950 and 813 Tg in 2001. A detailed analysis of ship emissions was made possible with the introduction of the Automatic Identification System (AIS) in 2004 and the subsequent availability of historical ship activity data. The first studies based on AIS data analysed air pollutants from the Port of Rotterdam in 2009 \cite{cotteleer_emissions_2011} and for the OSPAR II region in 2011 \cite{van_der_gon_methodologies_2010}. One of the best known models used to analyse ship emissions in the European context is the Ship Traffic Emission Assessment Model (STEAM), which is managed by the Finnish Meteorological Institute. STEAM has been used to determine emission quantities in the Baltic Sea \cite{jalkanen_comprehensive_2014}, in the Danish Straits \cite{jalkanen_extension_2012}, European Waters \cite{jalkanen_comprehensive_2016}, the Northern European ECA \cite{johansson_evolution_2013} and globally \cite{johansson_global_2017}. 

The Helmholtz-Zentrum Hereon is a leader in chemical transport modelling of pollutant emissions, as well as in the compilation of emission inventories, and has worked intensively on emissions from maritime transport \cite{karl_effects_2019} \cite{cnss_clean_2014} \cite{aulinger_impact_2016} \cite{matthias_impact_2016}. Hereon's Modular Ship Emission Model (MoSES) \cite{schwarzkopf_ship_2021} was used to compile a comprehensive maritime emission inventory for the North Sea and Baltic Sea region for 2015.

The study presented in this paper links maritime activity in the North Sea and Baltic Sea in 2015 to air pollutant emission quantities. These were calculated for all ships over 100 GT at 5-minute intervals resulting in the EUF (Europa University Flensburg) emission inventory, which is temporally and spatially highly-resolved. A bottom-up approach was used to consider nine leading air pollutants (CO\textsubscript{2}, NO\textsubscript{X}, PM\textsubscript{2.5}, SO\textsubscript{2} (sulphur dioxide), POA (primary organic aerosols), ash (mineral ash), CO (carbon monoxide), NMVOC (non-methane volatile organic compounds) and black (or elemental) carbon (BC)), as well as speed-dependent fuel and energy consumption. The inventory includes the emission from both main and auxiliary engines, as well as well-to-tank and tank-to-propeller emissions, and energy and fuel consumption values. The EUF inventory is available in CSV format and offers policy makers and non-modelling experts in particular an important starting point for their own transparent and reproducible analyses. Existing inventories covering large areas often have a maximum resolution of 1°x\,1° and a low temporal resolution, providing emissions as annual sums. In contrast, the EUF inventory provides higher resolutions and is therefore particularly suitable for the analysis of coastal areas or for use in chemical transport modelling. Furthermore, the inventory includes the pollutants PM, BC, CO and ash, which are not provided as standard in other inventories. 

Emission inventories are indispensable tools for environmental impact assessments and more generally for air pollution prevention measures through policy development and implementation. In addition, they can be used for the calculation of pollutant concentrations. The inventory presented here is particularly useful for the analysis of future emissions which take into account techno-economic and socio-ecological aspects (e.g. fuel switching, efficiency and sufficiency measures, and changes in trading volumes), and the analysis of energy requirements for shoreside power connections in port areas, which will become mandatory in Germany from 2023 \cite{bmwk_verordnung_2019}. 

\section*{Methods} 

All code and data sets (if not proprietary) are available on GitHub\cite{hilpert_klimaschiff_2022} and Zenodo\cite{hilpert_emission_2022}. 

\subsection*{Activity data}
The introduction of the AIS marked the beginning of digitisation in the shipping industry. Since the end of 2004, it has been mandatory for every ship over 100 GT to be equipped with an AIS-transmitter, which emits a signal every 6 seconds providing, for example, the IMO identification number (unique identifier), the position (longitude/latitude) and a time stamp. Companies such as MarineTraffic, Vesselfinder or IHS Fairplay collect and store these AIS signals and sell historical data. HELCOM (Baltic Marine Environmental Protection Commission) provides AIS data for the Baltic Sea region free of charge for research purposes. The presented emission inventory is based on high temporal resolution HELCOM \cite{helcom_emissions_2019} (Baltic Sea) and Vesselfinder \cite{vesselfinder_ais_2022} (North Sea) AIS data for 2015, covering the area between 65$^\circ$N, -5$^\circ$W, 48.3$^\circ$S and 30.7$^\circ$E, corresponding with the European Sulphur Emission Control Area (SECA). 

\subsection*{Ship routes}
Ship routes in the Baltic Sea and North Sea in 2015 were created from the acquired AIS data for all ships identified by their IMO number. Each route consists of segments that were formed from two consecutive AIS data points using the following steps:
\begin{itemize}
    \item Sorting of ship point data (longitude/latitude) by time stamp.
    \item Calculation of the distance between two consecutive time stamps using the Haversine equation.
    \item Calculation of the time duration between the two consecutive time stamps.
    \item Calculation of the speed of the ship between the time stamps based on time and distance.
    \item Resampling and interpolation to uniform 5 min time intervals (the spatial density of the received AIS signals on the open sea is much lower than in coast areas).
    \item Setting of ship speed to 0 knots if the calculated speed is < 1 knot. If the ship's speed is less than 1 knot, it is assumed that the ship is neither docking nor manoeuvring and thus only the auxiliary engine is active. This is confirmed by the analysis of the AIS Vesselfinder data, as the navigational status is on average (as of June 2015) 0.7 knots (at anchor) and 0.4 knots (moored) \cite{vesselfinder_movements_2022}. 
    \item Removal of all data points with a speed > 15 m/s (corresponds to approximately 29 knots), which may result as an artifact from erroneous longitude/latitude data. This happens, for example, if the distance between two points is particularly high, but the time difference is short, resulting in implausible calculated vessel speeds. If a vessel leaves the area under consideration, the route calculation cannot interpolate between two positions as usual since there may be several hours between them and an extended route outside the area under consideration. The route calculation is interrupted if a ship is at the edge of the area under consideration and the distance between the calculated points is > 300 m. The ship route calculation is restarted when the ship enters the area of consideration again.
\end{itemize}

\begin{figure}[ht!]
    \centering
    \includegraphics[scale=0.8]{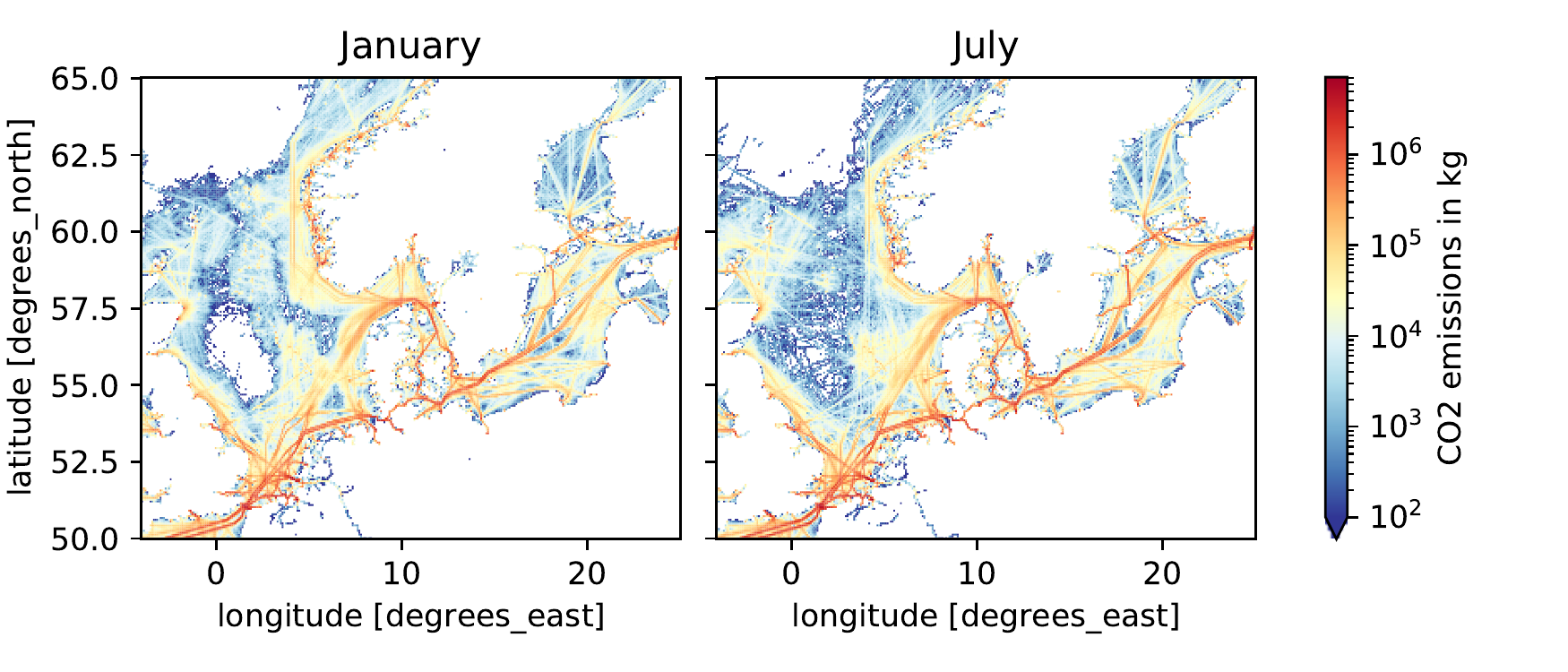}
    \caption{CO\textsubscript{2} emissions in kg in the North Sea and Baltic Sea for January and July, 2015.}
    \label{fig:co2_map_large}
\end{figure}

The initial results in Figure \ref{fig:co2_map_large} are a basic plausibility check for the bottom-up calculation of ship routes and emissions, as the known main shipping routes (cf. \cite{ec_baltic_2017}) are clearly identifiable from the AIS data–generated routes. Figure \ref{fig:co2_map_large} shows the annual gridded CO\textsubscript{2} emissions in the area under consideration. 

\subsection*{Ship types and characteristics}
\begin{figure}[ht]
    \centering
    \includegraphics[width=\textwidth
    ]{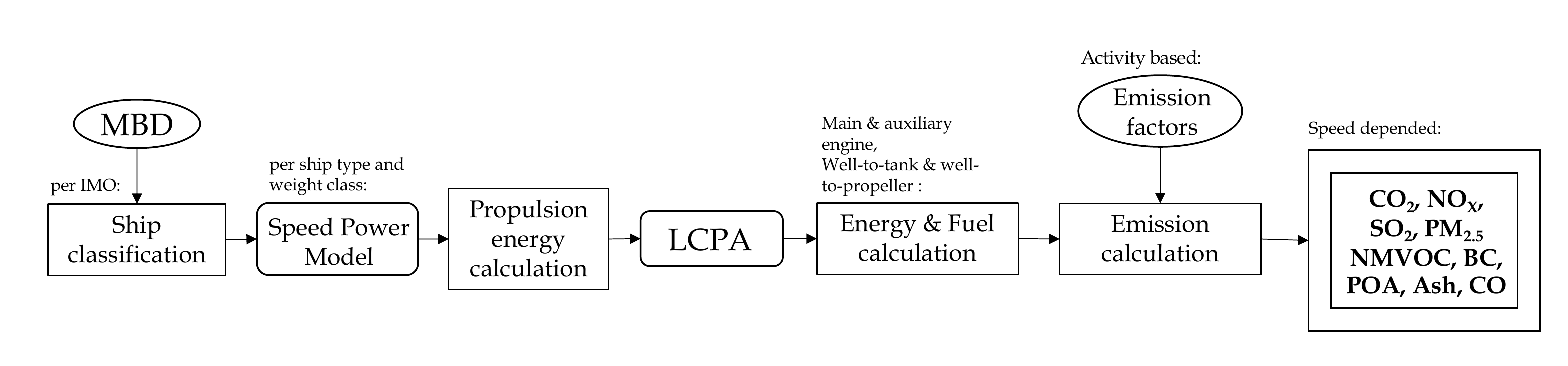}
    \caption{Schematic overview of methodological steps employed in the development of the EUF inventory with processing steps in squares. The squares with rounded corners contain the relevant input data for each step and the double square shows the final output.}
    \label{fig:schematic_overview}
\end{figure}

Decisive factors for the emissions from individual ships are primarily the ship type, size, age and engine configuration. \autoref{fig:schematic_overview} provides a schematic overview of the methodological steps to compile the inventory. 

 The Maritime Data Base (MDB) hosted by Vesselfinder, which includes 18,309 IMO identification numbers, was used cluster the ships using the listed \textit{AIS type}\cite{vesselfinder_ais_2022}\footnote{The MDB offers a range of information for each IMO identification number: Name, Built, Flag, Type (AIS-Type), Status, GT (Gross Tonnage), DWT (Deadweight tonnage), LOA (Length over all), LPP (Length between perpendiculars), Beam and Draught.}. Based on the AIS-ship types, standardised ship types and the expertise of shipbuilders\cite{nagel_shipbuilding_2020}, nine generic ship types were classified in this study: \textit{CarCarrier}, \textit{Container}, \textit{RoRo}, \textit{Bulker}, \textit{RoPax}, \textit{MPV} (Multi Purpose Vessel), \textit{Tanker}, \textit{Cruise}, and \textit{Diverse\footnote{The ship class \textit{Diverse} is mainly composed of 20\% tugs, 25\% fishing vessels and 16\% offshore supply/tugs. Tugs have a high installed power, but rarely use it due to high berthing times (70-80\%) and therefore are not defined as a separate class.}}. If the MDB did not specify the AIS type for a particular IMO identification number, the ship was assigned to the \textit{Diverse} ship type. A statistical analysis of the MDB data provided the weight distribution within each ship type. Based on international ship sizes (weight classes), between two and four weight classes were defined per ship type. Tanker and Container size classes are based on the international ship classes Handy Size (1), Handy Max (2), PanMax (3) and SuezMax (4). Table \ref{tab:weight_class} shows the set ship types by size classes (weight classes) (1-4). 

\begin{table}[!ht]
\caption{Ship characteristics in type and size (weight) classes 1-4 based on statistical analysis and ship building expertise \cite{nagel_shipbuilding_2020}.}
\label{tab:weight_class}
\begin{tabular}{l|c|c|c|cccc}
\toprule 
Type      & unit\textsuperscript{1} & main engine\textsuperscript{2} & auxiliary engine & 1 & 2   & 3  & 4  \\ 
\midrule
RoPax      & GT   & medium & medium & 0 - 24,999  &  from 25,000 &   &   \\
CarCarrier\textsuperscript{3} & GT   & slow & medium & 0 - 39.999 & from 40,000   &   &   \\
RoRo       & GT   & medium & medium & 0 - 24,999 & from 25,000   &   &   \\
Cruise     & GT   & medium & medium & 0 - 24,999   & from 25,000 &   &   \\
Diverse    & GT   & medium & medium & 0 - 1,999  & from 2,000    &   &   \\
Container\textsubscript{4}  & GT   & slow & medium & 0 - 17,499  & 17,500 - 54,999   & 55,000 - 144,999  & from 145,000\\
Tanker     & DWT  & slow & medium & 0 - 34,999  & 35,000 - 49,999   & 50,000 - 119,999  & from 120,000 \\
Bulker     & DWT  & slow & medium & 0 - 34,999   & 35,000 - 49,999  & 50,000 - 119,999  & from 120,000\\
MPV        & DWT  & slow & medium & 0 - 11,999 & from 12,000  &   &  \\
\bottomrule
\end{tabular}
\begin{tablenotes}
 \item \textsuperscript{1} Weight classes are measured in GT (gross tonnage) or DWT (dead weight tonnage), depending on the type of ship.
 \item \textsuperscript{2} Engine configuration differs between medium- or slow-speed diesel engines, depending on the type of ship. It is assumed that all ships within a particular class have the same engine configuration.
 \item \textsuperscript{3} Typical size specification for Pure Car Carriers (PCC) in terms of the number of transported cars was translated into GT.
 \item \textsuperscript{4} Typical sizes of TEU (Twenty Foot Equivalent Unit, standard for container ships) or PanMax, SuezMax etc. were translated into GT.
\end{tablenotes}
\end{table}

The ship types additionally differ in their engine configuration. A ship usually uses two sets of engines: the main engine(s), which provides the required propulsion power, and the auxiliary engine(s), which supply electrical power to the ship's electrical system via generators \cite{nagel_shipbuilding_2020}. Both set of engines run at different loads, have different engine characteristics and corresponding fuel consumption and associated emissions during operation. In general, a distinction is made between slow-speed, medium-speed and high-speed diesel engines \cite{dnv-gl_martime_2017}. The decisive factor here is that the engine speed with slow-speed engines operate from as low as 70 rpm up to 300 rpm, as is the case, for example, with most large two-stroke engines found on ships. Medium speed engines typically operate from approximately 300 to 900 rpm \cite{nagel_shipbuilding_2020}.

\subsection*{Energy and fuel consumption}
The required power for propulsion was calculated using a \textit{speed-power curve} (SPC) per ship type and size class \cite{nagel_shipbuilding_2020}, which determines the power consumption of each ship type as a function of speed. The calculation determines the required propeller power $P_D$\cite{hilpert_klimaschiff_2022}. The input variables are class-specific values such as average length, width and draught (arithmetic average of all ships with the set size class from the MDB). The SPC is simple to use and calibrate. It should be noted that the models represent generic, and thus average, ship types; the suitability of the application to model a single ship is dependent on the context and is not intended.

The auxiliary engine in all ships is assumed to be a medium-speed diesel engine. For simplification, it is assumed that the auxiliary diesel power is essentially constant over all speeds\cite{nagel_shipbuilding_2020}. Power values are based on EEDI specifications\cite{imo_fourth_2020}. Even though the power can increase at higher speeds, due to, for example, seawater cooling water or the operation of lube oil pumps, this effect should influence the overall results by less than 1\% \cite{nagel_shipbuilding_2020}.

For the calculation of expended energy and fuel consumption, a proprietary life cycle performance analysis (LCPA) tool was used. The tool was developed jointly by the Flensburger-Schiffsbaugesellschaft Gmbh (FSG), BALance Technology Consulting, SSPA Sweden, Teknologian tutkimskeskus VTT, IFEU, and Det Norske Veritas\cite{thiem_joules_2013},\cite{joules_lcpa_2014} to support the generation of ship designs. The LCPA tool and SPC model were coupled to determine the fuel mass flow, which is calculated using the assumed power of the engine together with the mechanical or electrical efficiencies, and the specific fuel consumption for a certain time period. The energy expended was derived from the amount of fuel consumed and its calorific value\cite{joules_lcpa_2014},\footnote{This step cannot be published, however can be traced and reproduced using the efficiency approximations of the ship's engines and the calculated propeller power within the SPC.}. It is assumed that all ships use low sulphur marine gas oil (LSMGO) (LHV (42.675 kJ/kg) with a sulphur content of less than 0.1\%, in line with IMO regulations within the European Sulphur Emission Control Area (SECA)\cite{imo_safety_2017}.

\subsection*{Emission calculation}
Tank-to-propeller pollutant emissions were determined using activity-based emission factors that relate to the fuel or energy consumption in combination with the navigational phase of the ship\cite{eea_emepeea_2019}. There are three navigation phases, as a distinction is made between underway, manoeuvring, and hotelling/at berth (manoeuvring and hotelling are often analysed as one).

The calculation of NO\textsubscript{x}, SO\textsubscript{2}, PM\textsubscript{2.5}, and CO\textsubscript{2} was performed within the LCPA model \cite{joules_lcpa_2014}. Due to the closed-source nature of the tool, only the qualitative calculation process is described. CO\textsubscript{2} emissions was calculated from the carbon fraction of the used fuel (LSMGO, 3,206 t\textsubscript{CO2}/t\textsubscript{fuel}) \cite{joules_lcpa_2014}. SO\textsubscript{2} emissions were determined using the acidification potential, based on the fuel sulphur content (0.1\%) and stoichiometric combustion to SO\textsubscript{2} in the energy converter \cite{joules_lcpa_2014}. PM emissions were calculated using the equation $\text{PM}=0.2+0.6\cdot S$ (g/kWh) for internal combustion engines \cite{imo_third_2015}\cite{joules_lcpa_2014}, where $S$ is the sulphur content of the fuel used. NO\textsubscript{x} emissions were assumed to be in line of the official MARPOL Tier regulations \cite{imo_third_2015}, depending on the ship construction date (see Table \ref{tab:tier_regulation}). The Tier I guideline was assumed for all ships built before 2000 \cite{imo_nitrogen_2020}, as this was closely set to the actual pollutant levels before 2000\cite{nagel_shipbuilding_2020}.

\begin{table}[!h]
\centering
\caption{Nitrogen oxide guidelines in the North Sea and Baltic Sea ECA (Emission Control Area) according to the IMO\cite{imo_third_2015}, MARPOL Annex VI.}
\label{tab:tier_regulation}
\begin{tabular}{p{1cm}p{5cm}p{3cm}p{3cm}p{3cm}}
\toprule 
& & \multicolumn{3}{c}{Total weighted cycle emission limit (g/kWh)}\\
& & \multicolumn{3}{c}{n = engine's rated speed (rpm)}\\
\hline
Tier    &   Ship construction date on or after & n<130   & n=130-1999 &  n>2000\\
\midrule
I   & 01. January 2000    & 17.0  & 45 $\cdot$ n\textsuperscript{(-0.2)}  & 9.8 \\
II  & 01. January 2011    & 14.4  & 44 $\cdot$ n\textsuperscript{(-0.2)}  & 7.7 \\
III & 01. January 2021    & 3.4   & 9  $\cdot$ n\textsuperscript{(-0.2)}  & 2.0 \\
\bottomrule
\end{tabular}
\end{table}

POA, NMVOC, Ash, BC and CO emissions were calculated using literature-based emission factors together with the fuel or energy consumption determined within the LCPA tool. A variety of emission factors for maritime applications can be found in the literature \cite{kristenen_energy_2015}, \cite{aulinger_impact_2016}, \cite{jalkanen_comprehensive_2016}, \cite{jun_co2_2001}, \cite{schwarzkopf_ship_2021}, \cite{eea_emepeea_2019}, \cite{eea_emission_2021}, \cite{entec_uk_limited_quantification_2002}. A key study is EMEP CORINAIR, the Guidance Document for Air Pollutant Emissions Inventory published by the European Environment Agency \cite{eea_emepcorinair_2007}. A 2019 version of the guide \cite{eea_emepeea_2019} gives activity-based emission factors for shipping (last update 12/2021\cite{eea_emission_2021}). All of the emission factors used, as well as an extensive literature review of additional emission factors for sensitivity analyses related to engine types, are available in the Zenodo repository \cite{hilpert_emission_2022}. 

Speed-related emissions $e(v)$ in kg were calculated from selected emission factors $ef$ and $E$ or $F$, the ship's energy and fuel consumption, respectively, using equations (1) or (2).
\begin{align}
   e (v) = ef \cdot E(v) \\
   e(v) = ef \cdot F(v)
\end{align}

In addition to the activity-based (tank-to-propeller) emissions, the life cycle emissions of the fuel used (well-to-tank) were also analysed. The LCPA tool provides well-to-tank emissions of NO\textsubscript{x}, SO\textsubscript{2}, PM\textsubscript{2.5} and CO\textsubscript{2} \cite{joules_lcpa_2014}. A full life cycle analysis can be used to evaluate the overall impact of a fuel in terms of greenhouse gas and pollutant emissions, including all phases from production to use. Typical well-to-tank life stages of marine fuels include extraction, transportation, conversion, carriage, and bunkering. The LCPA software used the UMBERTO\cite{ifu_hamburg_umberto_2022} life cycle software to determine the emission factors for well-to-tank emissions \cite{thiem_joules_2013}, \cite{joules_lcpa_2014}.

\section*{Data Records} 

All input and output data can be accessed via the Zenodo data repository\cite{hilpert_emission_2022}. The supporting model code is available on GitHub\cite{hilpert_klimaschiff_2022}. 

(1) The CSV file \textit{emission model} includes hourly emission quantities (well-to-tank and tank-to-propeller) and energy expended for the speed of the specific vessel type and size.

\begin{itemize}
    \item \textbf{Type} (column A): Ship type definition in three components separated by an underscore (\_): (1) the ship type (see Table \ref{tab:weight_class}), (2) the size class 1-4\cite{hilpert_emission_2022}) and (3) the age-dependent NO\textsubscript{x} regulation (Tier I, Tier II and FS (equivalent to Tier III)), e.g. ropax\_2\_tier 1 represents a RoPax vessel in size class 2 (above 25,000 GT), which was built before 01/01/2011.  
    \item \textbf{Engine} (column B): Propulsion (main) engine or Electric (Auxiliary engine) of a ship.
    \item \textbf{Speed [m/second]} (column C): The calculated speed over ground of a ship.
    \item \textbf{Energy (Well to tank) [J]} (column D): Energy expended for the production, transportation and distribution of the fuel used for propulsion.
    \item \textbf{\textit{Pollutant} (Well to tank) [kg]} (columns E-H): \textit{CO2}, \textit{SOx}, \textit{NOx} and \textit{PM} emissions during production, transportation and distribution of the fuel consummed (for SQ-2015 LSMGO). 
    \item \textbf{Energy [J]} (column I): Energy expended for the propulsion of the ship per speed.
    \item \textbf{Fuel Consumption [kg]} (column J): Tank-to-propeller fuel consumption.
    \item \textbf{\textit{Pollutant} [kg]} (columns K-S): Tank-to-propeller emission of \textit{CO2, SOx, NOx, PM, BC, ASH, POA, CO} and \textit{NMVOC}.
\end{itemize}

(2) The emission inventory \textit{ship\_emissions\_YYYYMMDD} is available as 396 csv files, one for each day in 2015 and December 2014. These contain the following data records.

\begin{itemize}
     \item \textbf{UniqueID} (column A): Unique identifier of ship. Due to the use of proprietary input data (AIS), the originally-used unique identifier (IMO-number) was replaced by a random number. 
     \item \textbf{Type} (column B): See the description for Type in the emission model input descriptor list above.
    \item \textbf{Datetime} (column C): Date and time stamp following the format YYYY-DD-MM HH:MM:SS.
    \item \textbf{Lat} (column D): Calculated latitudinal position of the ship in decimal degrees.
    \item \textbf{Lon} (column E): Calculated longitudinal position of the ship in decimal degrees.
    \item \textbf{Speed\_calc} (column F): Calculated speed in m/s from the calculated distance at a 5 min time interval .
    \item \textbf{Propulsion-Energy (Well to tank) [J]} (column G): Energy expended for the production, transportation and distribution of the fuel used in the main engine. 
    \item \textbf{Electrical-Energy (Well to tank) [J]} (column H): Energy expended for the production, transportation and distribution of the fuel used in the auxiliary engine. 
    \item \textbf{Propulsion-\textit{Pollutant} (Well to tank) [kg]} (columns I, K, M and O): \textit{CO2}, \textit{SOx}, \textit{NOx} and \textit{PM} (PM\textsubscript{2.5}) emissions from the production, transportation and distribution of the fuel needed by the main engine for propulsion in the respective time interval.
    \item \textbf{Electrical-\textit{Pollutant} (Well to tank) [kg]} (columns J, L, N and P): \textit{CO2}, \textit{SOx}, \textit{NOx} and \textit{PM} (PM\textsubscript{2.5}) emissions from the production, transportation and distribution of the fuel needed by the auxiliary engine in the respective time interval.
    \item \textbf{Propulsion-Energy [J]} (column Q): Energy content of the fuel used for propulsion (tank-to-propeller) of main engine in the respective time interval. 
    \item \textbf{Electrical-Energy [J]} (column R): Energy content of the fuel used for the auxiliary engine (tank-to-propeller in the respective time interval. 
    \item \textbf{Propulsion-Fuel Consumption [kg]} (column S): Fuel consumption for the propulsion (main engine) of the vessel in the respective time interval. 
    \item \textbf{Electrical-Fuel Consumption [kg]} (column T): Fuel consumption for the hotelling load (auxiliary engine) of the vessel in the respective time interval. 
    \item \textbf{Propulsion-\textit{Pollutant} [kg]} (columns U, W, Y, AA, AC, AE, AG, AI and AK): \textit{CO2, SOx, NOx, PM, BC, ASH, POA, CO, NMVOC} emissions from the main engine during operation (tank-to-propeller) of the ship. 
    \item \textbf{Electrical-\textit{Pollutant} [kg]} (columns V, X, Z, AB, AD, AF, AH, AJ and AL): \textit{CO2, SOx, NOx, PM, BC, ASH, POA, CO, NMVOC} emissions from the auxiliary engine during operation (tank-to-propeller) of the ship. 
\end{itemize}

(3) In the interest of transparency and reproducibility, a number of supporting data sets are available on Zenodo\cite{hilpert_emission_2022}. The \textit{speed-power-model.xlsx} file summarises the assumed power (kW) for the main and auxiliary engines as well as inputs for the SPC calculation (length (m), width (m), draught (m), the shape-dependent variable c\textsubscript{b}, the propulsion efficiency and the wetted surface of the ship in m\textsuperscript{2}) specific to speed, ship type and weight class. The \textit{emission\_factors.xlsx} file summarises the emission factors specific to engine type, navigation phase and pollutant as well as an extensive, referenced list of factors for sensitivity analyses. The \textit{analysis.xlsx} file contains additional information for further analyses of, for example, emission quantities as assessed within different models for cross-model comparisons. Also included is the time spent at a particular speed, as well as the number of ships considered in each ship and weight class. The \textit{supplementary\_material.pdf} file summarises regulations and targets for maritime emissions\cite{imo_third_2015}\cite{imo_fourth_2020}. This file also contains the equations to calculate the propulsion power $P_D$.\\

(4) All related code including static input data is available on GitHub\cite{hilpert_klimaschiff_2022}. Within the folder \textit{emission\_model}, the  \textit{maximum\_speed\_per\_type.csv} file lists the maximum possible speed in m/s \cite{nagel_shipbuilding_2020}. The \textit{ship\_weightclass\_mapper.csv} file assigns the different ship types and sizes to weight types (DWT or GT) and the lower and upper bounds of the weight classes connecting Tier I, II and FS with the year of construction and the typical engine rpm for each specification. Both \textit{ship\_type\_fsg\_mdb\_mapper.xlsx} file and \textit{short\_long\_name\_mapper.xlsx} file are used for the pre-processing of ship type classification. Within the folder \textit{lcpa-models} are \textit{SensitivityAnalysis-ShipeTypeName-Rev2.csv} files, which are hourly emission models for all ship types and sizes. The \textit{scr} folder contains the model code itself, pre-processing steps and routines for result and input data analyses. 


\section*{Technical Validation}

\subsection*{Comparison with existing emission inventories.}
Crucial in the creation of an emissions inventory through bottom-up modelling is the validation of the results obtained. There is no possibility to compare the chosen assumptions and related results with actual emission measurements. Initial studies on measuring ship emissions using drones are underway \cite{winnes_-board_2016}, exemplified by the SCIPPER project \cite{scipper_scipper_2021} and efforts by the European Maritime Safety Agency (EMSA). However, only individual ships on a specific voyage can be measured. A cross-model comparison, however, allows a statement about the quality of the obtained results. Useful in this context is Helmholtz-Zentrum Hereon's model MoSES \cite{schwarzkopf_ship_2021} and the Finnish Meteorological Institute's Ship Traffic Emissions Assessment Model (STEAM) \cite{jalkanen_comprehensive_2016}. The two models have been used to create emission inventories for the North Sea and Baltic Sea Area: MoSES\cite{schwarzkopf_ship_2021} for 2015, STEAM - in this application - for 2011\cite{jalkanen_comprehensive_2016}. Table \ref{tab:emission_comparison_euf_steam_moses} summarises the emission quantities, base year and number of ships analysed for the three aforementioned inventories.

\begin{table}[!ht]
    \centering
            \caption{Result values from MoSES \cite{schwarzkopf_ship_2021}, STEAM \cite{jalkanen_comprehensive_2016} and the EUF inventory in Gg per year and respective deviations to the EUF results in \%.}
    \label{tab:emission_comparison_euf_steam_moses}
    \begin{tabular}{l|lll|ll}
    \toprule
        ~ & MoSES & EUF & STEAM & Diff MoSES (\%) & Diff STEAM (\%) \\ \hline
        Year & 2015 & 2015 & 2011 & & \\
        \hline
        Number of ships & 21,845 & 16,632 & n/a & 23.86 & n/a \\
    \midrule
        SO\textsubscript{2} & 32.55 & 24.90 & 192.10 & 30.70 & 671.34 \\ 
        NO\textsubscript{x} & 897.97 & 818.00 & 806.20 & 9.78 & -1.44 \\ 
        BC & 13.89 & 0.44 & n/a & 3,036.76 & n/a \\ 
        POA & 17.96 & 11.62 & n/a & 54.50 & n/a \\ 
        MA & 0.32 & 0.22 & n/a & 46.85 & n/a \\ 
        CO\textsubscript{2} & 44,886.43 & 34,931.98 & 35,740.00 & 28.50 & 2.31 \\ 
        CO & 38.31 & 38.04 & 57.30 & 0.72 & 50.64 \\ 
        PM\textsubscript{2.5} & 29.83 & 13.94 & 38.30 & 113.98 & 174.74 \\ 
        NMVOC & 11.12 & 17.32 & n/a & -35.81 & n/a \\ 
    \bottomrule
    \end{tabular}
\end{table}

In general, a good agreement can be observed between the values from EUF, STEAM and MoSES for most pollutants. The emission quantities from the STEAM and MoSES inventories tend to be higher than those from the EUF model. This can be explained, among other things, by the number of ships investigated in the area under consideration. Comparing the EUF and STEAM values, it can be seen that there is a significant deviation in SO\textsubscript{x} emissions. This is largely due to the introduction of the 0.1\% sulphur guideline in 2015, which can also partially explain the deviation in PM\textsubscript{2.5} emission results, as the reduced sulphur content in the fuel also reduces particulate matter emission. Further comparative analyses are possible based on \cite{hilpert_emission_2022}.

\begin{figure}[!ht]
    \centering
    \includegraphics[width=0.68\textwidth]{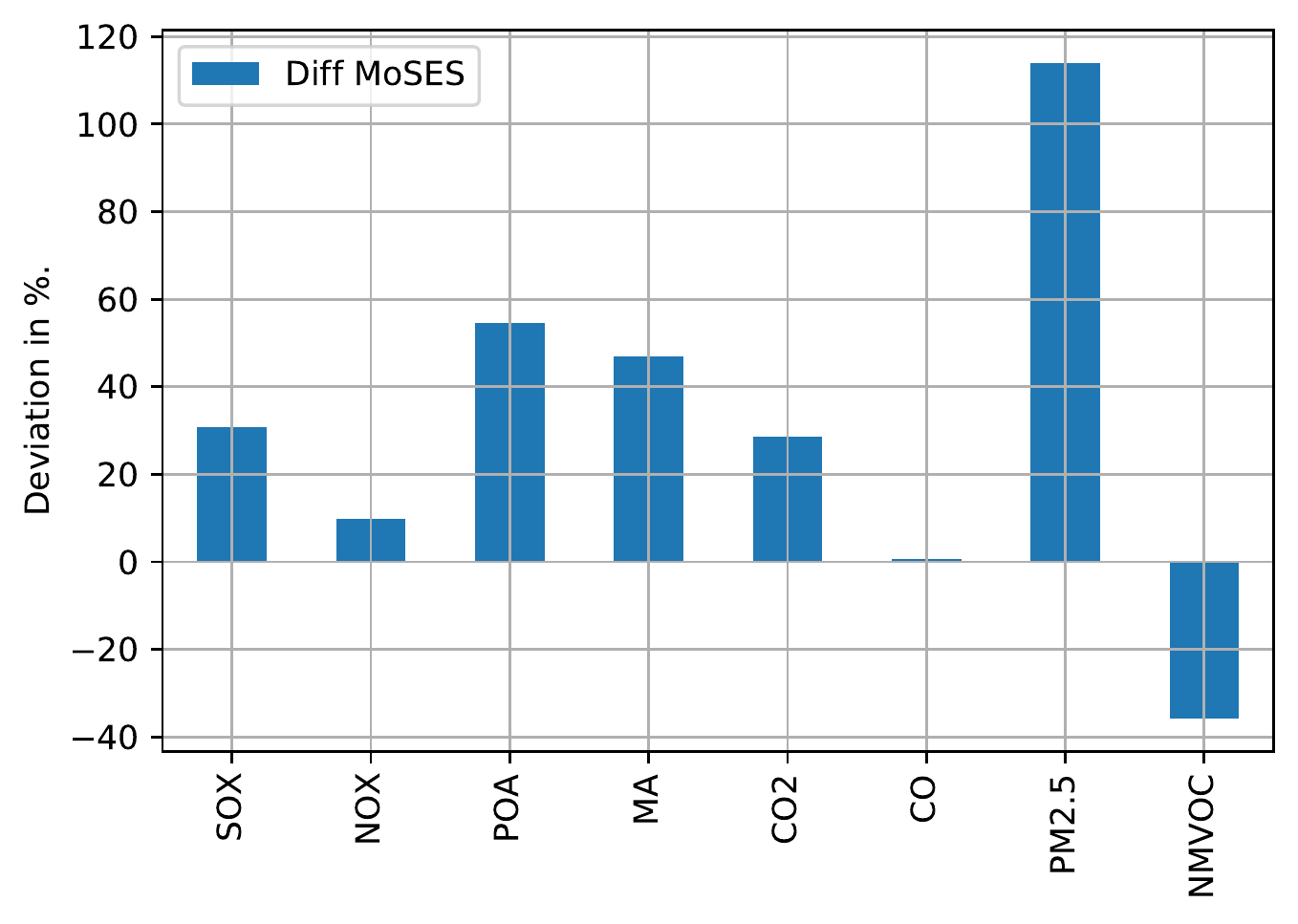}
    \caption{Deviation (in \%) of values (emission quantity in the EUF inventory) to MoSES \cite{schwarzkopf_ship_2021} for SQ-2015. MA = Mineral Ash = Ash}
    \label{fig:result_comparison_steam_moses_euf}
\end{figure}

Figure \ref{fig:result_comparison_steam_moses_euf} provides a detailed comparison of emissions from the EUF and MoSES\cite{schwarzkopf_ship_2021} inventory and the deviations of the results of MoSES from the EUF emission inventory in percent. There is a tendency for MoSES emission values to be higher than those in the EUF inventory, which may be due to the number of ships considered and the different calculation approaches for energy consumption and related emissions. A comprehensive comparison of the selected fuel and energy modelling is difficult due to the different methodological paths chosen. However, CO\textsubscript{2} emissions are a stable indicator, since the carbon from the required fuel burns almost entirely to CO\textsubscript{2}. The variation here is consistent with the variations in the number of ships considered. The SO\textsubscript{2} emissions are in the same order of magnitude in the MoSeS and EUF values. The differences in NO\textsubscript{x}, POA, and Ash are within a reasonable range and can be justified by the different emission factors used. The differences in NMVOC and BC are striking. Contrary to the general trend, NMVOC emissions are higher in the EUF inventory. BC emissions, on the other hand, are an order of magnitude lower in the EUF emission inventory. 

The EUF model uses mainly fuel-based emission factors, while the emission factors within MoSES are related to the energy expended. MoSES uses a factor of 0.03 g/kWh for BC emissions from the main engine and 0.15 g/kWh for the auxiliary engine, as well as a decided adjustment of the factor to low loads of the engine. The EUF model uses the European Energy Agency's \cite{eea_emission_2021} factors (see Table \ref{tab:emission_factors_euf_moses}). A more in-depth analysis that examines variances arising from energy expenditure is recommended. 

\begin{table}[h]
    \centering
        \caption{Emission factors used in MoSES \cite{schwarzkopf_ship_2021} and EUF, FSC = fuel sulphur content (0.1\,\%) and SFOC = specific fuel oil consumption.}
    \label{tab:emission_factors_euf_moses}
    \begin{tabular}{l|lllll}
    \toprule
     Model & BC & POA & Ash & CO & NMVOC \\
   \midrule
     MoSES & 0.03 g/kWh & 0.2 g/kWh & FSC $\cdot$ SFOC $\cdot$ 0.02 g/kWh & 0.54 g/kWh & 0.5 g/kWh\\
     EUF & 0.0329 kg/t & 0.09 g/kWh & 0.0002 kg/t & 3.47 kg/t & 1.52 kg/t\\
     \bottomrule
    \end{tabular}
\end{table}

The analysis of the different values from STEAM, MoSES and EUF show how crucial the choice of emission factors is. The factors for the EUF model were chosen based on an extensive literature review; particular emphasis was placed on the timeliness of the source. Nevertheless, there are large uncertainties that must be taken into account when interpreting the values from all models.

\subsection*{Uncertainties.} 
The results of the study must be viewed critically with regard to the methods and models used. The results are discussed concerning (1) (input) data (quality) and (2) the general modelling approach and connected assumptions. 

(1) Uncertainties in the reliability of the emission inventories arise from the use of the closed-source LCPA model and the AIS data, which are proprietary. This limits the reproducibility and transparency of the values generated. The creation of ship routes is influenced by the quality of the AIS data used and by the interpolation routines chosen. The AIS data used has a high data ranking (A or B are most accurate \cite{vesselfinder_ais_2022}), which makes the generated values more reliable. Since the end of 2018, satellite-processed AIS data are available, which should increase the accuracy of results. The LCPA model is a closed-source model. However it builds on the expertise of ship builders and environmental analysts. The emission values and associated calculation steps used with the LCPA model can be traced with sufficient accuracy.

(2) Within the general study approach, the chosen speed reduction criteria may also affect the results. Unlike other studies such as Schwarzkopf \textit{et al.}\cite{schwarzkopf_ship_2021}, this study does not consider engine load for the calculation of fuel and power consumption. The speed-power curves take this into account, but the auxiliary engine is only partially responsive to the engine load and is simplified for modelling purposes. Unlike STEAM \cite{jalkanen_extension_2012}, \cite{jalkanen_comprehensive_2014}, the influence of waves and currents is not directly considered in the EUF model. However, the SPCs tend to have high power assumptions, so consideration of waves and currents was not considered essential. The loading condition is not mentioned in any of the analysed studies\cite{jalkanen_comprehensive_2014}, \cite{schwarzkopf_ship_2021} and is not considered in the EUF analysis. 

Generic ship types have been defined, based on a complementary data set obtained from Vesselfinder, which can be compared with the AIS types \cite{vesselfinder_ais_2022}. Nevertheless, the integration of a more accurate ship type assignment system could further refine the emission values. Additionally, only ships over 100 GT were utilised in the EUF model so the study does not provide an assessment of all emissions. 

For future scenarios, fleet developments have to be taken into account. In 2017, maritime transport accounted for nearly 90\% of international trade measured in tonne-kilometers \cite{dnv-gl_martime_2017}. Maritime trade volumes have tripled since 1980. The DNV expects seaborne trade to be 35\% higher in 2030 than in 2017 and to increase by a further 12\% until 2040. The UNCTAD\cite{unctad_review_2019} projected an average annual growth of 3.5\% between 2019 and 2024, with the largest relative growth expected in the gas and container cargo sectors, each increasing by 135-150\,\%. Bulk carriers are projected to increase by 40\% by 2050, based on a combination of an increase in non-coal bulk trade and an eventual reduction in coal shipments \cite{unctad_review_2019}. Looking specifically at the future development of container ships and their contribution to total emissions, it can be assumed that future emissions will increase disproportionately to the increase in goods transported. 

\subsection*{Further use of data}
The data in the EUF emission inventory can be used to assess air pollution regulations and incorporate different emission control or propulsion technologies. The data can also be used in chemical transport modelling to evaluate the resulting air pollutant concentration from maritime emissions. Subsequently, gridded population data could be used together with the results from chemical transport modelling to estimate human health impacts of air pollutants. Because the EUF inventory also provides fuel and energy consumption at a high resolution, new emission factors can be applied for different pollutants and the inventory can be expanded accordingly. The high spatial resolution of the data set makes it possible to analyse the shore-to-ship power demand for renewable energy to reduce emissions from the auxiliary engine at berth.

\section*{Usage Notes} 
The emission inventory and all supporting data sets, with the exception of the AIS data, are available under an open license on Zenodo\cite{hilpert_emission_2022}. The data can be processed using Excel or any other data processing software capable of processing CSV files, such as Python or R. 

AIS data for the Baltic Sea can be obtained from HELCOM via a data agreement. The AIS data for the North Sea is proprietary data, which can be purchased from Vesselfinder. The LCPA model used is also proprietary, but the associated computational pathways are described qualitatively, allowing sufficient traceability of the results. 

\section*{Code availability} 

The code for the underlying calculations and post-processing functions are available under the BSD 3-Clause licence \cite{hilpert_klimaschiff_2022}. The analysis uses Python as a programming language.

\bibliography{manuscript}


\section*{Acknowledgements} 
The authors would like to thank the team from Helmholtz-Zentrum Hereon, Daniel Schwarzkopf, Jan Arndt , Volker Matthias and Armin Aulinger for their constructive support and feed-back. The authors also thank Frank Borasch from Flensburger Schiffsbau-Gesellschaft Nobiskrug. A special thanks goes to Jonathan Mole for his tireless and highly-skilled linguistic proofreading work. 

\section*{Author contributions statement}
F.D. led the project, collected and assembled the data, was responsible for the research design, analysed the results and prepared the manuscript, S.H. developed the model code, analysed the results and reviewed the manuscript. R.N. developed the SPC model and reviewed the manuscript. R.P. reviewed the manuscript and supported the formal analysis. 

\section*{Competing interests}
This research was supported by the EKSH (Society for Energy and Climate Protection Schleswig-Holstein). No known financial gain for the funding giver is recognised.




\end{document}